\documentclass[3p,onecolumn,times,number]{elsarticle}
\usepackage{amsmath}
\usepackage{amssymb}
\usepackage{graphics}
\usepackage{multirow}
\usepackage{graphicx}
\usepackage{epsfig}
\usepackage{dcolumn}
\usepackage{bm}
\usepackage{lineno}
\linenumbers 

\begin{document}
\begin{frontmatter}

\title{Optimization of various isolation techniques to develop low noise, radiation hard double-sided silicon strip detectors for the CBM Silicon Tracking System }
\author[label1]{S.~Chatterji\corref{cor}}
\ead{S.Chatterji@gsi.de}
\author[label2]{M.~Singla}
\author[label1]{W.F.J.~M\"{u}ller}
\author[label1]{J.M.~Heuser}

\cortext[cor]{Corresponding author}

\address[label1]{GSI Helmholtzzentrum f\"{u}r Schwerionenforschung GmbH, Darmstadt, Germany}
\address[label2]{Goethe University, Frankfurt, Germany}

\begin{abstract}
This paper reports on the design optimization done for Double Sided silicon microStrip Detectors~(DSSDs) to reduce the Equivalent Noise Charge (ENC) and to maximize the breakdown voltage and Charge Collection Efficiency. Various isolation techniques have been explored and a detailed comparison has been studied to optimize the detector performance.  For the evaluation of the performance of the silicon detectors, a radiation damage model has been included. The neutron fluence is expected to be 2$\times$10$^{13}$~n$_{eq}$cm$^{-2}$ per year for five years of expected CBM run with intermediate periods of warm maintenance, cold maintenance and shutdown. Transient simulations have been performed to estimate the charge collection performance of the irradiated detectors and simulations have been verified with experimental data. \end{abstract}
\begin{keyword}
Double Sided silicon Strip Detector, Equivalent Noise Charge~(ENC), radiation hardness, Transient simulation, Charge collection, Strip isolation, TCAD, SYNOPSYS

\end{keyword}
\end{frontmatter}

\section{Introduction}
\label{}

The mission of the Compressed Baryonic Matter~(CBM) ~\cite{aa} experiment at the Facility for Antiproton and Ion Research~(FAIR) now under construction in Darmstadt, Germany, is to explore the QCD phase diagram in the region of high baryon densities. The layout of CBM detectors is driven by the corresponding experimental requirements concerning material budget, reaction rates, radiation tolerance, particle densities and selectivity. The detector will face the problem of measuring Au + Au interactions at 25~GeV/nucleon and up to 10 MHz rate producing up to 1000 charged particle tracks per event. The core of the CBM detector is Silicon Tracking Station (STS) located in a large-aperture dipole magnet. 
The STS of the CBM will consist of eight tracking stations placed at a distance between 10-100~cm downstream of the target in a dipole magnet with internal field of 1~Tm. Each station is a modular structure of DSSDs of different sizes to match the non-uniform channel occupancy distribution from the beam pipe to the periphery. The sensors will be held by low-mass carbon fibre support structures with read-out electronics outside of the detector aperture, thus minimizing the Coulomb scattering of particles. 

\begin{table*}[htb!] \small
\centering
\caption{Fluence profile of neutrons expected for CBM STS}
\begin{tabular}{|c|c|c|c|c|c|c|}
\hline\hline
Year & Fluence (n$_{eq}$ cm$^{-2}$) & N$_{eff}$ (cm$^{-3}$)& $\tau_{e}$ (ns) & $\tau_{h}$ (ns) & V$_{fd}$ (V) & V$_{op}$ (V)\\ [0.5ex]
\hline
1 & 2$\times$10$^{13}$ &  2.80$\times$10$^{11}$  & 1140 & 1050 & 28 & 70   \\ \hline
2 & 4$\times$10$^{13}$ & -1.54$\times$10$^{11}$ & 570 & 527  &20 &50   \\ \hline
3 & 6$\times$10$^{13}$ & -5.35$\times$10$^{11}$ & 380 & 351 & 44& 110 \\\hline
4 & 8$\times$10$^{13}$& -8.84$\times$10$^{11}$ & 285 & 263 & 75& 150 \\\hline
5 & 1$\times$10$^{14}$& -12.1$\times$10$^{11}$  & 228 & 211 & 100 & 200 \\[0.5ex]
\hline
\end{tabular}
\label{table:table1}
\end{table*}

We aim to develop low-noise radiation hard DSSDs for the CBM STS. This is a challenging task keeping in mind that CBM uses fast self-triggering electronics; the shaping time of the front end chip used for early prototyping ~\cite{n-XYTER} is 19~ns (fast shaper) and 140~ns (slow shaper) respectively. The dominant contribution to the total Equivalent Noise Charge (ENC) is given by the series capacitive and the resistive components ~\cite{bb,cc,dd}. These two noise components are further inversely proportional to the shaping time of the preamplifier. The other less significant contributions to the total ENC like shot noise has also been taken into account in this study. Parameters relevant for the extraction of these two dominant components of ENC like interstrip capacitance and metal trace resistance has been studied. The various contributions to the total Equivalent Noise Charge (ENC) are given as follows:\\
(a) Series Capacitive Noise 
\begin{equation}
\label{eqn1}
a + b.C_{tot}        
\end{equation}

(b) Series Resistive Noise
\begin{equation}
\label{eqn2}
24.C_{tot}(pF).\sqrt{R_{s}(\Omega)/\tau_{s}(ns)} e^{-}
\end{equation}

(c)Shot Noise
\begin{equation}
\label{eqn3}
108.\sqrt{I_{leak}(\mu A).\tau_{s}(ns)} e^{-}
\end{equation}

(d)Parallel Resistive Noise
\begin{equation}
\label{eqn4}
 24.\sqrt{\tau_{s}(ns)/R_{p}(M\Omega)} e^{-}
\end{equation}

where a and b are electrical parameters dependent on the front end electronics, C$_{tot}$ is the total capacitance, R$_{s}$ is the series resistance, $\tau_{s}$ is the shaping time of the preamplifier, I$_{leak}$ is the leakage current and R$_{p}$ corresponds to the bias resistor value.

Table~\ref{table:table1} shows the expected neutron fluence for five years of expected CBM runtime. The maximum fluence is expected to be 1$\times$10$^{14}$~n$_{eq}$cm$^{-2}$ which is similar to the Large Hadron Collider~(LHC) radiation environment. One can observe a deterioration of carrier life time with fluence which will have an impact on the Charge Collection Efficiency~(CCE), especially on the p-side since this side collects less mobile holes.
In order to understand radiation damage, some of the CBM prototype DSSDs having dimension 1.6~cm$\times$1.6~cm and resistivity 6~k$\Omega$cm were irradiated at the KRI cyclotron facility in St. Petersburg, Russia. These detectors were measured just after irradiation without any periods of annealing. The variation of leakage current and interstrip resistance with neutron fluence has been measured. The radiation damage model implemented in TCAD simulations is able to reproduce the measured observations. 

In order to investigate the life time of DSSDs, it is imperative to extract CCE as a function of fluence for which one has to understand strip isolation in particular on the ohmic side. Hence we are exploring various isolation techniques, for example P-stop, P-Spray, Modulated P-spray (conventional isolation techniques) and also a new isolation technique namely, Schottky barrier. TCAD simulations have been performed to optimize the total ENC, strip isolation and to maximize the breakdown voltage (V$_{bd}$) and CCE. Transient simulations can be used to extract the operating voltage of DSSDs which in turn depends on the isolation technique. 

\begin{figure}[h]
\centering
\includegraphics*[height=70mm]{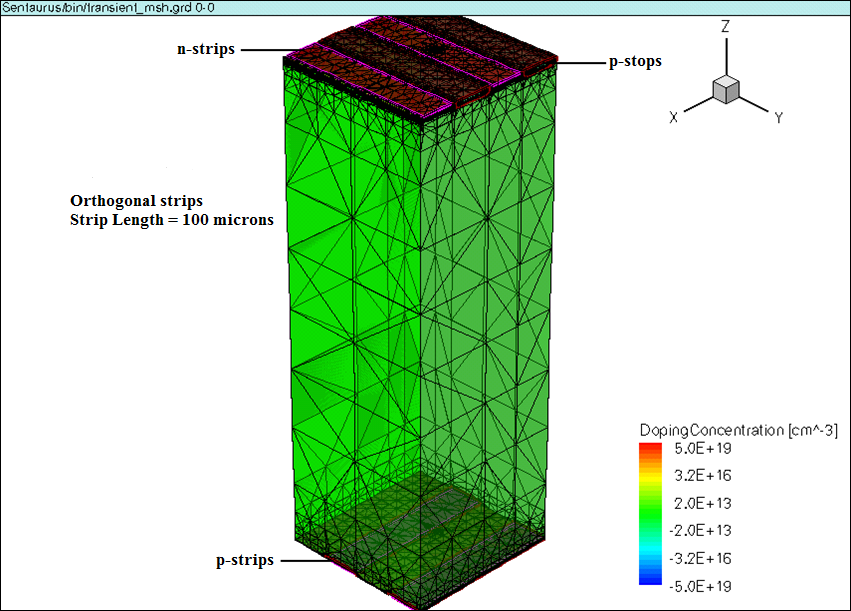}
\caption{Slice of 3-dimensional DSSD grid having orthogonal strips.}
\label{fig1}
\end{figure}

\begin{figure}[h]
\vspace*{4mm}
\centering
\includegraphics*[height=100mm]{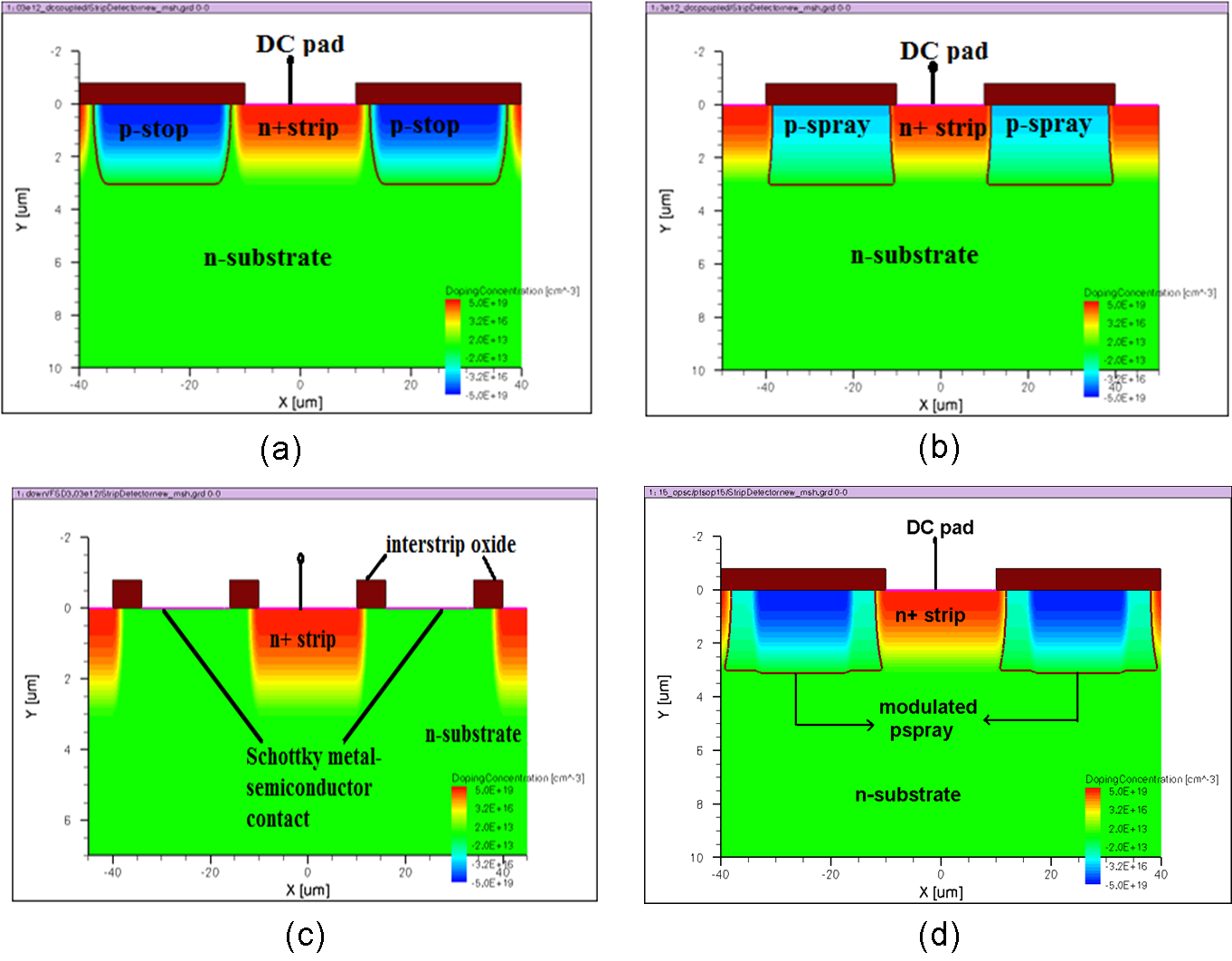}
\caption{Simulated TCAD grid of DSSD having four different isolation techniques (a) P-stop; (b) P-spray; (c) Schottky barrier; (d) Modulated P-spray.}
\label{fig2}
\end{figure}

\section{Structure of simulated devices}
\subsection{Simulated structure of DSSD}
\label{}

The AC-coupled DSSDs with n-type silicon substrate of 300~$\mu$m thickness have a strip width of 20~$\mu$m, a pitch of 50~$\mu$m with orthogonal strips, as shown in Figure~\ref{fig1}. The n-strips are orthogonal to the p-strips. We are using moderate resistivity n-type silicon of around 5-6~k$\Omega$cm corresponding to effective doping concentration of 9$\times$10$^{11}$~cm$^{-3}$  and 7.5$\times$10$^{11}$~cm$^{-3}$, respectively. All p$^{+}$ and n$^{+}$ implants were approximated by assuming a Gaussian profile with a peak concentration of 5$\times$10$^{19}$~cm$^{-3}$ at the surface. It is assumed that the lateral diffusion depth at the junction curvature of the implants is equal to 0.8 times the vertical junction depth (X$_{j}$). Depletion is attained by applying a positive voltage to the ohmic side and grounding the junction side. The thickness of the coupling oxide taken is around 200~nm while the interstrip gap is filled with a thicker oxide of around 800~nm. The surface oxide charge (Q$_{f}$) has been taken to be 3$\times$10$^{11}$~cm$^{-2}$ before irradiation for a good quality oxide. For the $<111>$ silicon orientation used in detector fabrication, the amount of charge is expected to increase and saturate at 3$\times$10$^{12}$~cm$^{-2}$, under heavy irradiation of the order of a few hundred krad ~\cite{ee,ff}. The conventional isolation techniques namely P-stop, P-spray and Modulated P-spray have been compared with a new isolation technique called Schottky barrier. The TCAD simulation grids having these isolation techniques can be seen in Figure ~\ref{fig2}(a-d). The conventional isolation techniques have been defined by assuming a Gaussian profile with a peak concentration of 5$\times$10$^{19}$~cm$^{-3}$ at the surface for P-stop and 4$\times$10$^{16}$~cm$^{-3}$ (low dose) for P-spray. For modulated P-spray, an optimization study has been done for various combinations of P-stop width and P-spray dose which will be discussed later in this paper. The new isolation technique namely Schottky barrier can be defined either through metal work function value or through barrier height which in turn depends on the substrate type and the metal used for schottky contact. For Aluminium, the barrier height is 0.72~eV for n-type silicon while for p-type silicon, the barrier height is 0.58~eV ~\cite{gg,zz}.\\
For accurate simulations, it is imperative to use correct boundary conditions. The reason being that we are simulating only a part of a full device, hence we need to be sure that this does not affect the accuracy of the simulation. One possible way to do this is to simulate larger area of the device and then to look at the results only in the central region, away from the boundaries. The other approach is to simulate some repeating unit of device and then to make sure that the boundary conditions are appropriate. The default is to use reflecting (Neumann) boundary conditions for a device with transitional symmetry and mirror symmetry. However in our device with orthogonal strips, there is transitional symmetry in x and y direction but no mirror symmetry. Hence we have applied periodic boundary conditions (PBC) in x and y direction for device having orthogonal strips.

\subsection{Models implemented in TCAD simulation}
\label{}
SYNOPSYS~\cite{yy}, a finite element semiconductor simulation package was used to determine the electrical behaviour of these devices.
The effective doping concentration (N$_{eff}$) is parameterized using Hamburg model ~\cite{hh,ii} which consist of three components, a short term beneficial annealing component, a stable damage part and a reverse annealing component. 
To incorporate the effect of increase in leakage current with fluence, the minority carrier lifetime $\tau$ has been changed in our simulation package using the definition of Kraner ~\cite{jj} as follows:
\begin{equation}
1/\tau = 1/\tau_{0} + \beta.\phi_{eq}
\end{equation}

where $\tau_{0}$  is the minority carrier lifetime of the initial wafer, $\phi_{eq}$ is the integrated fluence, and $\beta$ is the trapping time. The default value of $\tau_{0}$ for the electrons and holes is 10~$\mu$s  and 3~$\mu$s in Sentaurus Device. These values have been modified in Scharfetter relation ~\cite{kk} to a value of 1~ms for electrons and 0.3~ms for holes as is expected for detector grade silicon. 

When high-energy particles such as hadrons pass through a detector, they collide with silicon atoms and displace them from their lattice sites, resulting in pairs of interstitial atoms and vacancies. These defects may recombine, or they may form complexes with each other or with existing impurities in the silicon ~\cite{ll,xx,ww}. These defects introduce extra energy levels within the bandgap of the silicon. SYNOPSYS simulates this bulk radiation damage by directly modelling the dynamics of these traps ~\cite{mm}. So, the user has to provide SYNOPSYS with the concentrations and parameters of the traps. The trap model used here is based on the work done at the University of Perugia ~\cite{nn}.

\begin{table*}[htb!] \small
\centering
\caption{Comparison of measured and simulated data for irradiated DSSDs}
\begin{tabular}{|c|c|c|c|c|}
\hline\hline
Fluence n$_{eq}$ cm$^{-2}$ & \multicolumn{2}{c|} {Leakage Current @ 20$^{0}$C (nA)} & \multicolumn{2}{c|} {V$_{fd}$ (V)}\\ [0.5ex]
\cline{2-3}\cline{4-5}
 & Measured & Simulated &Measured & Simulated  \\[0.5ex]\hline
2.06$\times$10$^{12}$ & 5.4$\times$10$^{3}$ &  5.68$\times$10$^{3}$ & 47$ \pm $5 & 45  \\ \hline
3.03$\times$10$^{12}$ & 9.5$\times$10$^{3}$ &  8.31$\times$10$^{3}$ & 29$\pm $5 & 37  \\ \hline
3.93$\times$10$^{12}$ & 10.7$\times$10$^{3}$ &  10.7$\times$10$^{3}$ & 29 $\pm$ 5 & 32 \\\hline
11.20$\times$10$^{12}$& 35.1$\times$10$^{3}$ &  30.3$\times$10$^{3}$ & 9 $\pm$ 5 & 9 \\\hline
20.60$\times$10$^{12}$& 66.5$\times$10$^{3}$ &  58.8$\times$10$^{3}$ & 50 $\pm$ 5 & 48 \\[0.5ex]
\hline
\end{tabular}
\label{table:table2}
\end{table*}

\begin{table*}[htb!] \small
\centering
\caption{Expected power consumption of irradiated DSSDs}
\begin{tabular}{|c|c|c|c|c|c|c|}
\hline\hline
Fluence & \multicolumn{2}{c|} {V$_{op}$(V)} & Leakage Current &  \multicolumn{2}{c|} {R$_{int}$@ V$_{op}$ (G$\Omega$)}  & Power Consumption\\ [0.5ex]
\cline{2-3}\cline{5-6}
n$_{eq}$ cm$^{-2}$& & & @ -10$^{0}$C(nA) & & &  V$_{op}$ $\times$ I$_{leak}$ ($\mu$W mm$^{-1}$) \\ [0.5ex]

 & Measured & Simulated & &Measured & Simulated &  \\[0.5ex]\hline
3.03$\times$10$^{12}$ &60$ \pm$ 5 & 55 & 346 & 0.2$\pm$ 0.05  & 1 &8.10$\times$10$^{-2}$  \\ \hline
3.93$\times$10$^{12}$ &64 $\pm$ 5 & 62 & 448 & 0.1$\pm$ 0.05  & 1.5 &11.20$\times$10$^{-2}$ \\\hline
11.20$\times$10$^{12}$&8 $\pm$ 2 & 20 & 1250 & 0.1$\pm$ 0.05  & 0.6 &3.90$\times$10$^{-2}$ \\\hline
20.60$\times$10$^{12}$&120 $\pm$ 10 & 105 & 2420 & 0.1$\pm$ 0.05  & 0.2 &113.44$\times$10$^{-2}$ \\[0.5ex]
\hline
\end{tabular}
\label{table:table3}
\end{table*}

\section{Results and discussion}
\subsection{Leakage current and depletion behaviour of sensors}
\label{}

\begin{figure}[h]
\centering
\includegraphics*[height=60mm]{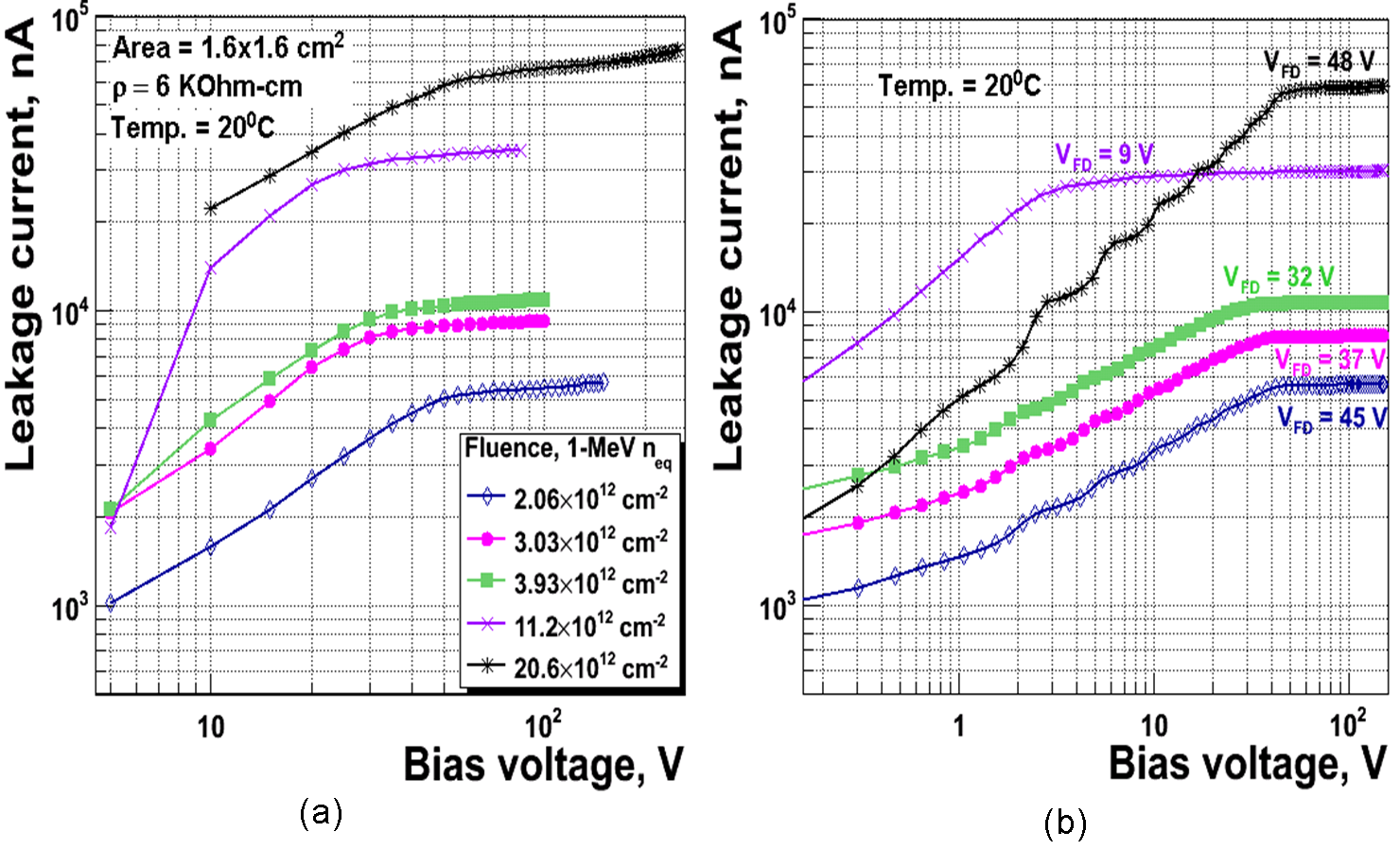}
\caption{Current-Voltage characterstics of irradiated DSSDs having P-stop isolation obtained through (a) measurements and (b) simulations.}
\label{fig3}
\end{figure}

\begin{figure}[h]
\vspace*{4mm}
\centering
\includegraphics*[height=55mm]{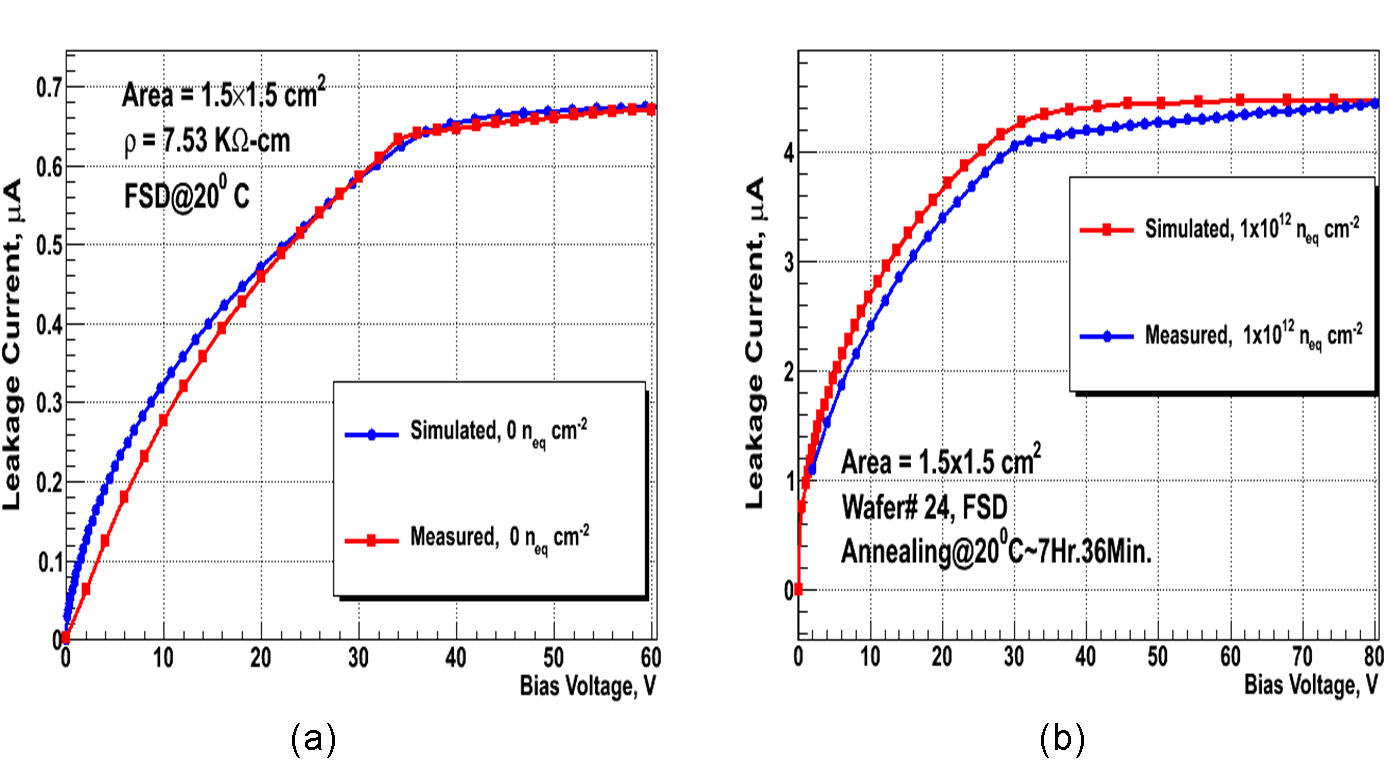}
\caption{Comparison of measured variation of leakage current vs. bias voltage with simulation for DSSDs having schottky barrier isolation (a) at Zero fluence and (b) after irradiation with 1$\times$10$^{12}$~n$_{eq}$cm$^{-2}$.  }
\label{fig4}
\end{figure}

Current-Voltage~(I-V) characteristics were simulated for DSSDs and compared with measurements. In DSSDs, most of the bulk leakage current is due to electron-hole pairs being produced by Shockley-Read-Hall (SRH) generation ~\cite{mm} in the depletion region, which are then swept to the electrodes by the electric field. Some of the CBM prototype DSSDs have been irradiated with neutrons at KRI cyclotron facility in St. Petersburg, Russia. The irradiated data are shown in Figure~\ref{fig3}(a). The full depletion voltage (V$_{fd}$) indicates that the type inversion occurs at around 11.20$\times$10$^{12}$~n$_{eq}$cm$^{-2}$. The full depletion voltage decreases with increasing fluence until the point of type inversion and after type inversion it starts increasing with fluence. The simulated I-V curve shown in Figure~\ref{fig3}(b) matches qualitatively with the measured curves for the irradiated sensors. A comparison of measured parameters with the simulated values is shown in Table~\ref{table:table2}. The leakage current increases with fluence, thus increasing the shot noise component in the ENC calculations. However this can be controlled by operating the sensors at cryogenic temperatures. The extracted damage constant ($\alpha$) from simulation is about 3.73$\times$10$^{-17}$~A cm$^{-1}$ while from measurement is about 4.2$\times$10$^{-17}$~A cm$^{-1}$. A similar agreement has also been found for DSSDs having Schottky barrier both for unirradiated and for the irradiated ones including the effect of annealing as can be seen in the figures~\ref{fig4}(a) and ~\ref{fig4}(b).

\subsection{Series Capacitance and resistance of sensors }
\label{}
\begin{figure}[h]
\vspace*{4mm}
\centering
\includegraphics*[height=60mm]{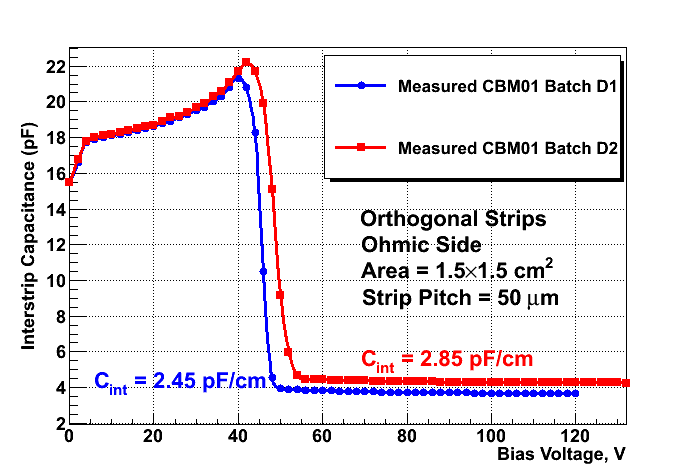}
\caption{Measured variation of C$_{int}$ with bias voltage for DSSDs having P-spray isolation.}
\label{fig5}
\end{figure}

\begin{figure}[h]
\vspace*{4mm}
\centering
\includegraphics*[height=60mm]{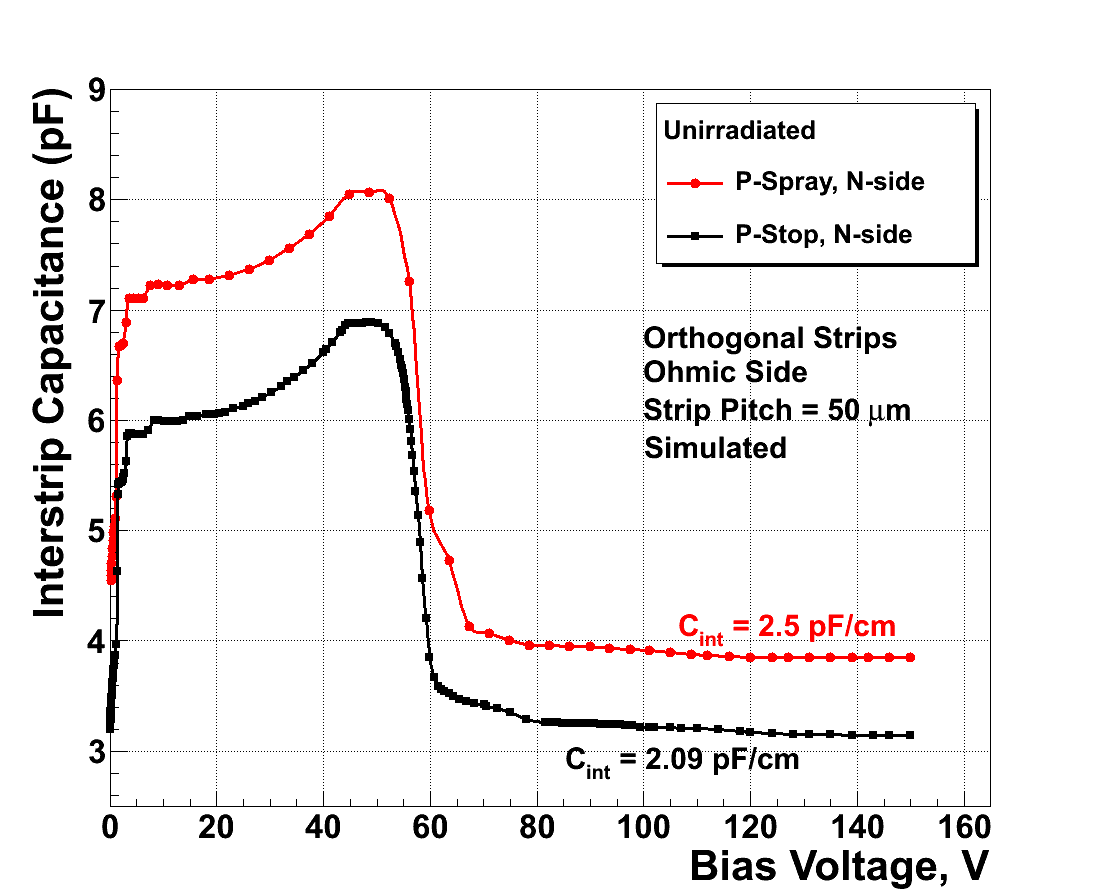}
\caption{Simulated variation of C$_{int}$ with bias voltage for DSSDs having P-spray and P-stop isolation.}
\label{fig6}
\end{figure}

Series capacitance and series resistance (trace resistance) are the dominant factors contributing to the total ENC. The strip capacitance is the sum of the capacitance to the backplane and direct capacitance to the adjacent strips (interstrip capacitance). For small pitch microstrip detectors having       300~$\mu$m thickness, the interstrip capacitance (C$_{int}$) is expected to dominate over the backplane capacitance. We have used TCAD simulations to extract the C$_{int}$ both before and after irradiation. Figure~\ref{fig5} shows the measured variation of ohmic side C$_{int}$ versus bias voltage (V$_{bias}$) for non-irradiated DSSDs having P-spray isolation. Figure~\ref{fig6} shows the simulated variation of C$_{int}$ versus V$_{bias}$ for both P-stop and P-spray isolation. For both the cases, it has been found that C$_{int}$ initially increases till the point of full depletion. After full depletion, there is a steep fall in C$_{int}$ and then it saturates.

\begin{figure}[h]
\vspace*{2mm}
\centering
\includegraphics*[height=60mm]{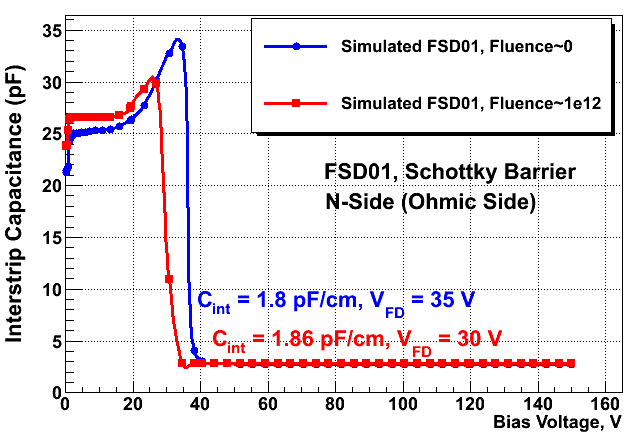}
\caption{Simulated variation of C$_{int}$ with bias voltage for DSSDs having Schottky barrier isolation.}
\label{fig7}
\end{figure}

\begin{figure}[htb!]
\vspace*{2mm}
\centering
\includegraphics*[height=65mm]{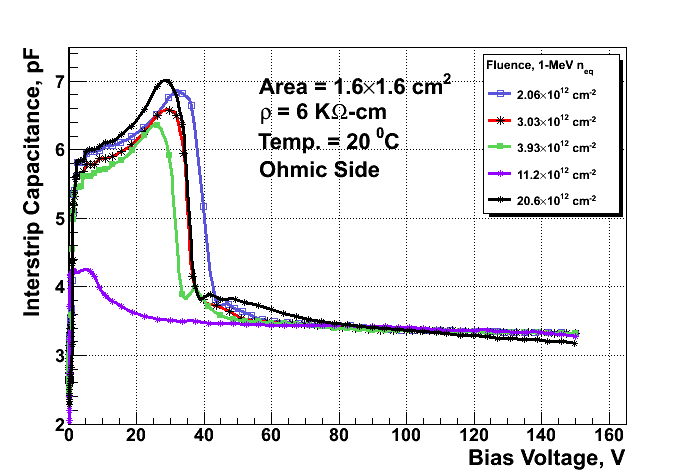}
\caption{Interstrip capacitance vs. bias voltage for irradiated DSSDs equipped with P-stop isolation technique.}
\label{fig8}
\end{figure}

\begin{figure}[htb!]
\vspace*{-1mm}
\centering
\includegraphics*[height=60mm]{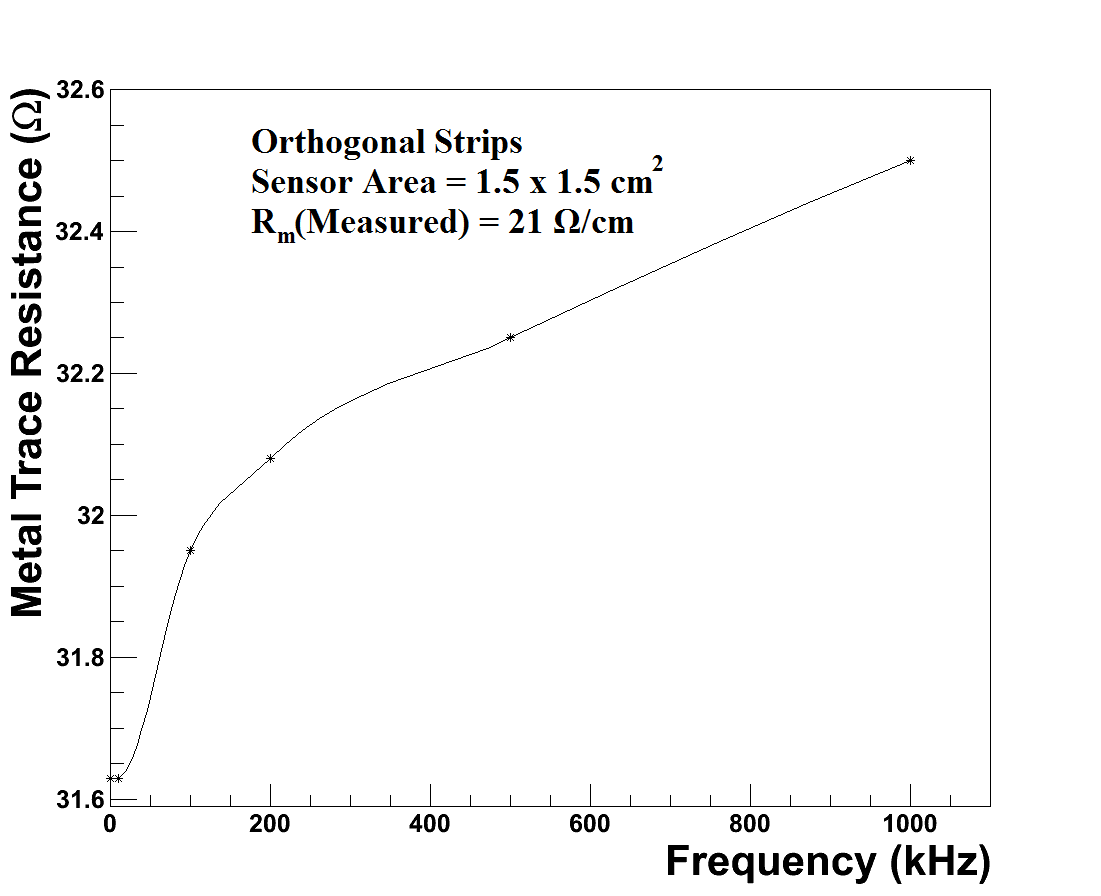}
\caption{Measured trace resistance vs. frequency for DSSD equipped with P-spray isolation technique.}
\label{fig9}
\end{figure}

One can notice that there is a nice match between simulated and measured C$_{int}$ especially after full depletion. Also it becomes clear that P-stop isolation gives lower capacitance as compared with P-spray isolation. We have simulated the expected variation of C$_{int}$ versus V$_{bias}$ for Schottky barrier isolation in order to compare it with the conventional isolation techniques. It is clear from Figure~\ref{fig7} that C$_{int}$ for Schottky barrier isolation is slightly lower as compared with P-stop isolation technique after full depletion. However before full depletion, C$_{int}$ for Schottky barrier isolation is too high both for unirradiated as also for irradiated DSSDs which could be detrimental if the DSSDs have to be operated under-depleted at high fluences.
We wanted to study if there is an impact of irradiation on the interstrip capacitance. For this purpose, we have used our radiation damage model which has already been validated with measurements. Figure~\ref{fig8} shows the simulated variation of ohmic side C$_{int}$ versus V$_{bias}$ after irradiation for DSSDs equipped with P-stop. The shape of the C$_{int}$ remains the same as for non-irradiated case. We do not observe a drastic change in the value of C$_{int}$ with irradiation. It can be noted that the value of C$_{int}$ before full depletion initially decreases with fluence till the point of type-inversion and increase thereafter. Figure~\ref{fig9} shows the variation of metal trace resistance as a function of frequency. The measured resistance has been found to be around 21~$\Omega$cm$^{-1}$. The contribution from the parallel resistance to the ENC is negligible since it is inversely proportional to the value of bias resistor whose value is of order of few M~$\Omega$'s.

\subsection{Strip isolation and the charge collection efficiency}
\label{}

\begin{figure}[htb!]
\vspace*{2mm}
\centering
\includegraphics*[height=60mm]{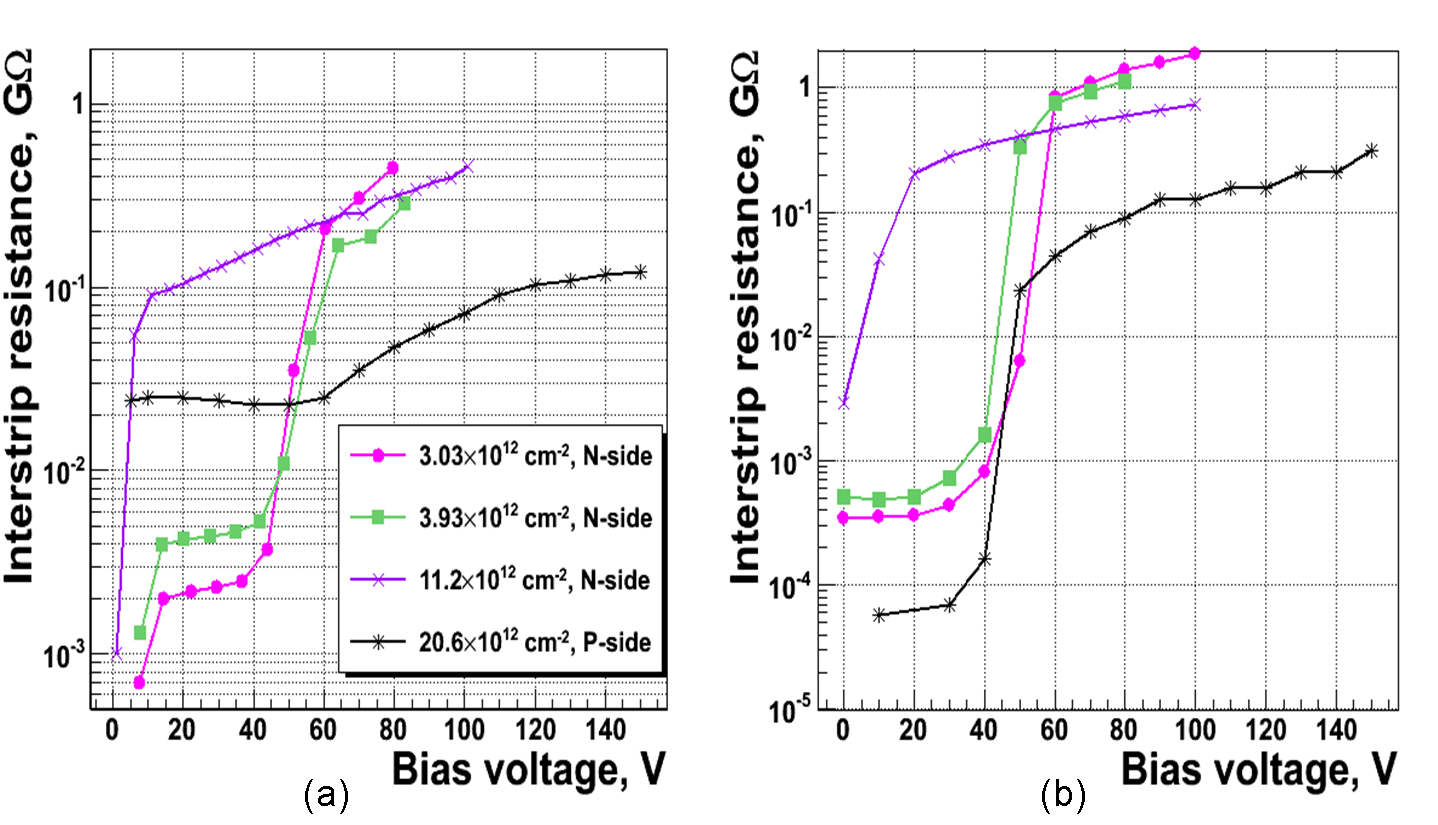}
\caption{Variation of interstrip resistance vs. bias voltage of irradiated DSSDs having P-stop isolation technique obtained through (a) measurements and (b) simulations.}
\label{fig10}
\end{figure}

\begin{figure}[htb!]
\vspace*{4mm}
\centering
\includegraphics*[height=50mm]{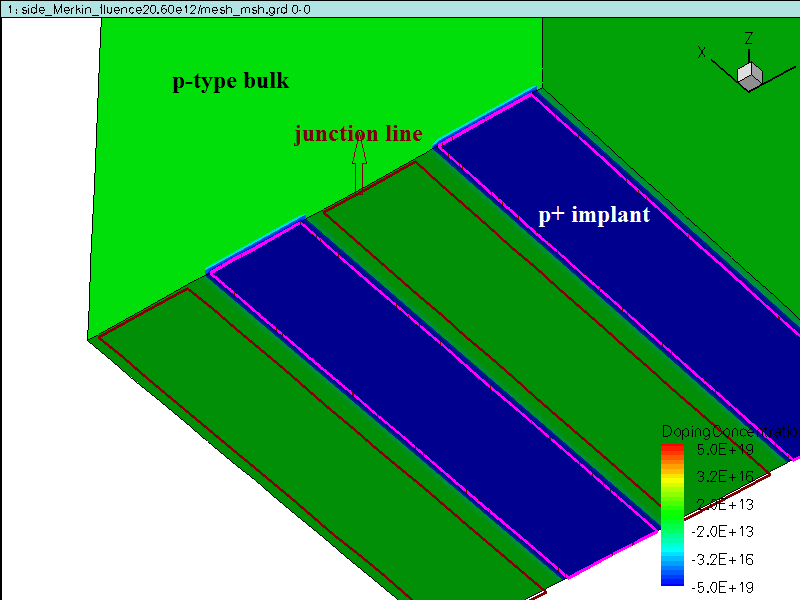}
\caption{Junction line formed by type-inverted bulk and e$^{-}$ accumulation layer on the ohmic side.}
\label{fig11}
\end{figure}

\begin{figure}[htb!]
\vspace*{4mm}
\centering
\includegraphics*[height=50mm]{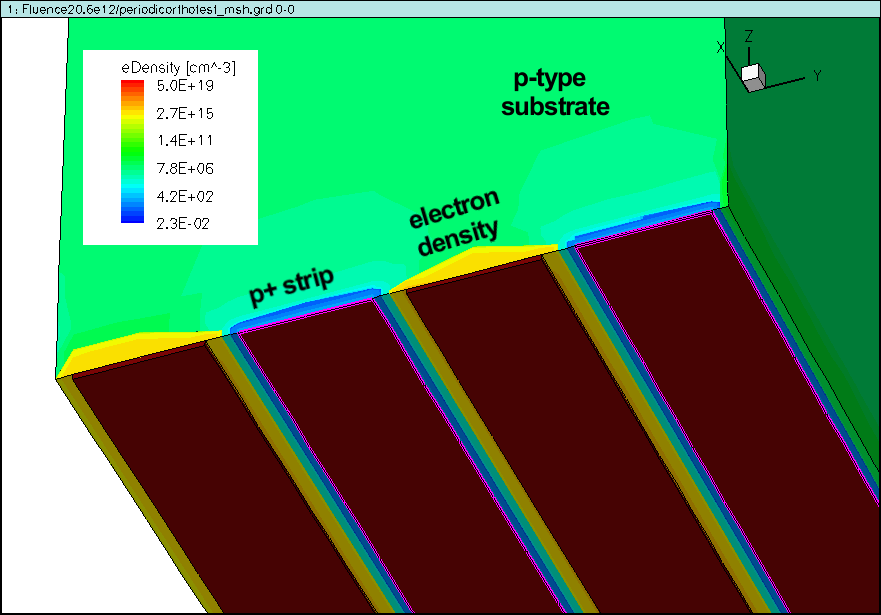}
\caption{Electron density between the P-strips in type-inverted DSSD.}
\label{fig12}
\end{figure}

In order to investigate the life time of DSSDs, it is imperative to extract CCE as a function of fluence. One can understand strip isolation by studying interstrip resistance (R$_{int}$) in particular on the ohmic side. In figure~\ref{fig10}(a), the measured variation of ohmic side R$_{int}$ with V$_{bias}$ can be seen for different fluences. For the fluences up to type inversion, interstrip resistance is very low before full depletion and it increases steeply at full depletion and continues to increase slightly thereafter. However for type-inverted sensors, interstrip resistance is of the order of tens of M$\Omega$'s even before full depletion and saturates at around 100~M$\Omega$ after the operating voltage is reached. In type inverted sensors, this effect can be attributed to electron accumulation layer ~\cite{oo} due to which the type of silicon between the p-strips remains n-type as can be seen in Figure~\ref{fig11}. To confirm this effect, e$^{-}$ density between the p-strips has been extracted as shown in Figure~\ref{fig12}. The high e$^{-}$ density between p-strips confirm the presence of n-type substrate between the p-strips even after type-inversion. Thus, isolation on the p side is realized even at low biases after type inversion. Figure~\ref{fig10}(b) shows the simulated values of interstrip resistance for the irradiated sensors for the same fluences as in measurements. A good match has been found for the fluences up to type inversion. However there is some discrepancy with the type inverted case for biases below full depletion. The simulated trap model is not able to reproduce the effect of electron accumulation layer in R$_{int}$ simulation. Though this effect is reflected accurately in the transient simulations as discussed below.  Also it should be noted from Figure~\ref{fig10}(b) that the operating voltage is around 1.5-2.0 times the V$_{fd}$ depending on the fluence. It is advisable to reduce the operating voltage since it directly increases the power consumption. Table~\ref{table:table3} shows the expected power consumption of DSSDs irradiated up to 2.06$\times$10$^{13}$~n$_{eq}$cm$^{-2}$. One can observe that the power consumption increases steeply after type-inversion. 

\begin{figure}[htb!]
\vspace*{4mm}
\centering
\includegraphics*[height=60mm]{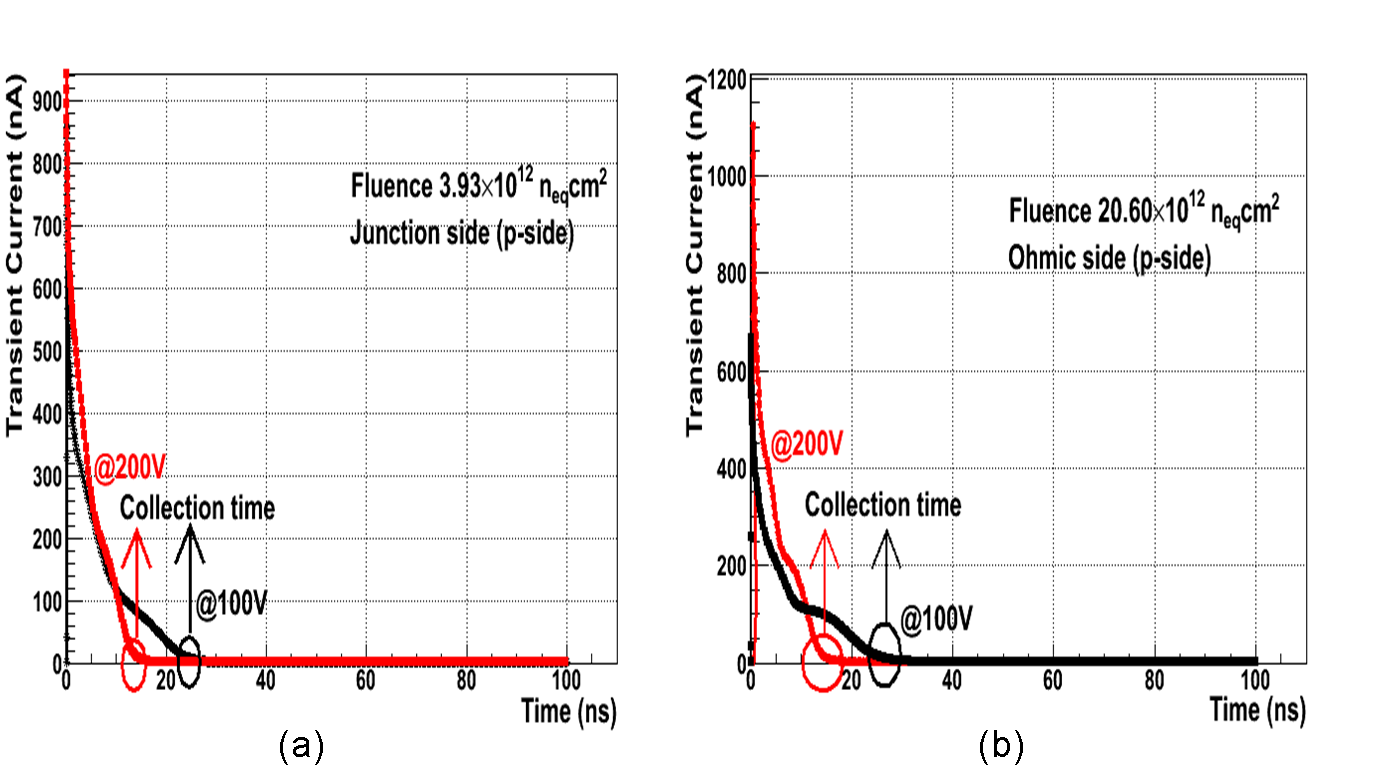}
\caption{Transient signal on the p-side at V$_{bias}$ 100V \& 200V for DSSDs irradiated to (a) 3.93$\times$10$^{12}$~n$_{eq}$cm$^{-2}$ and (b) 20.60$\times$10$^{12}$~n$_{eq}$cm$^{-2}$. }
\label{fig13}
\end{figure}

\begin{figure}[htb!]
\vspace*{4mm}
\centering
\includegraphics*[height=60mm]{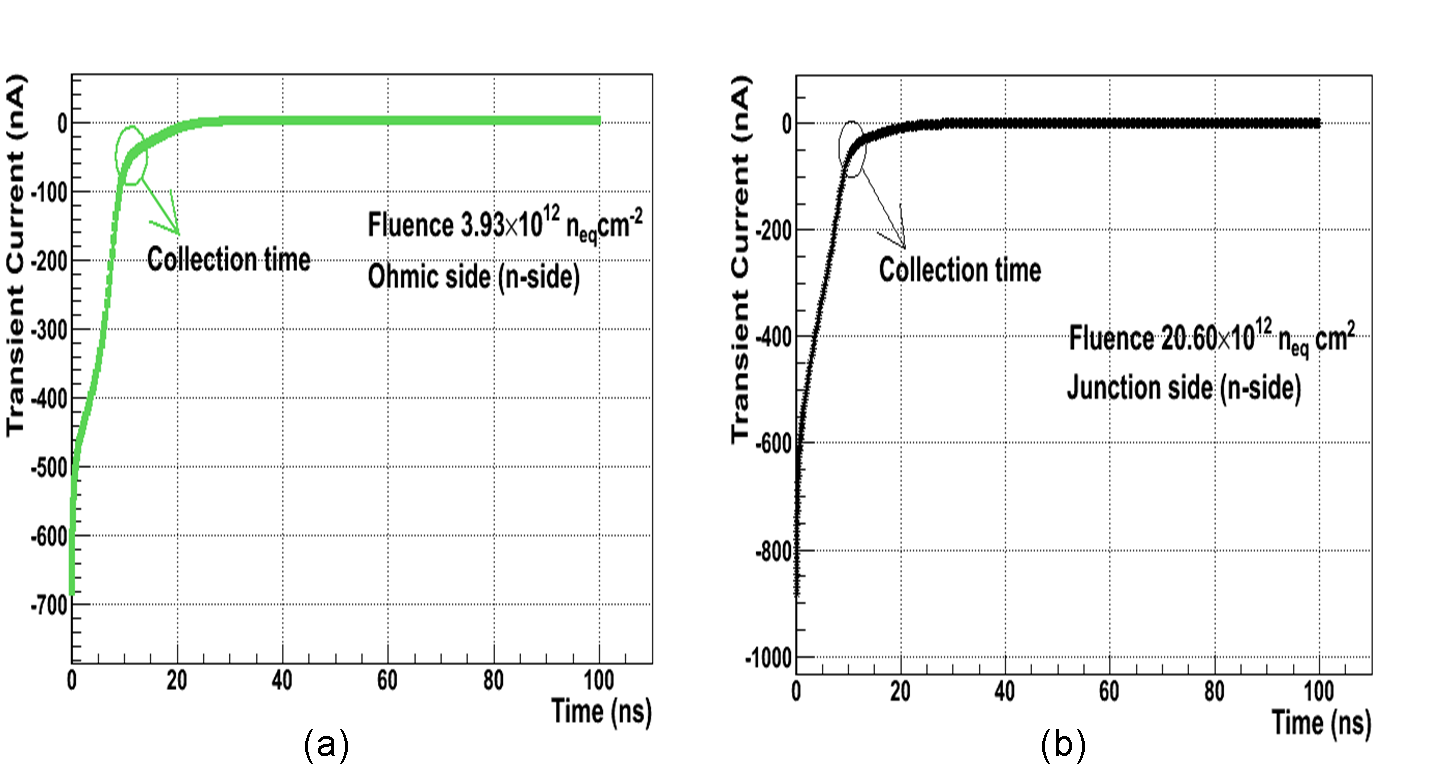}
\caption{Transient signal on the n-side at V$_{bias}$ 100V for DSSD irradiated to (a) 3.93$\times$10$^{12}$~n$_{eq}$cm$^{-2}$ and (b) 20.60$\times$10$^{12}$~n$_{eq}$cm$^{-2}$.}
\label{fig14}
\end{figure}

\begin{figure}[htb!]
\vspace*{4mm}
\centering
\includegraphics*[height=60mm]{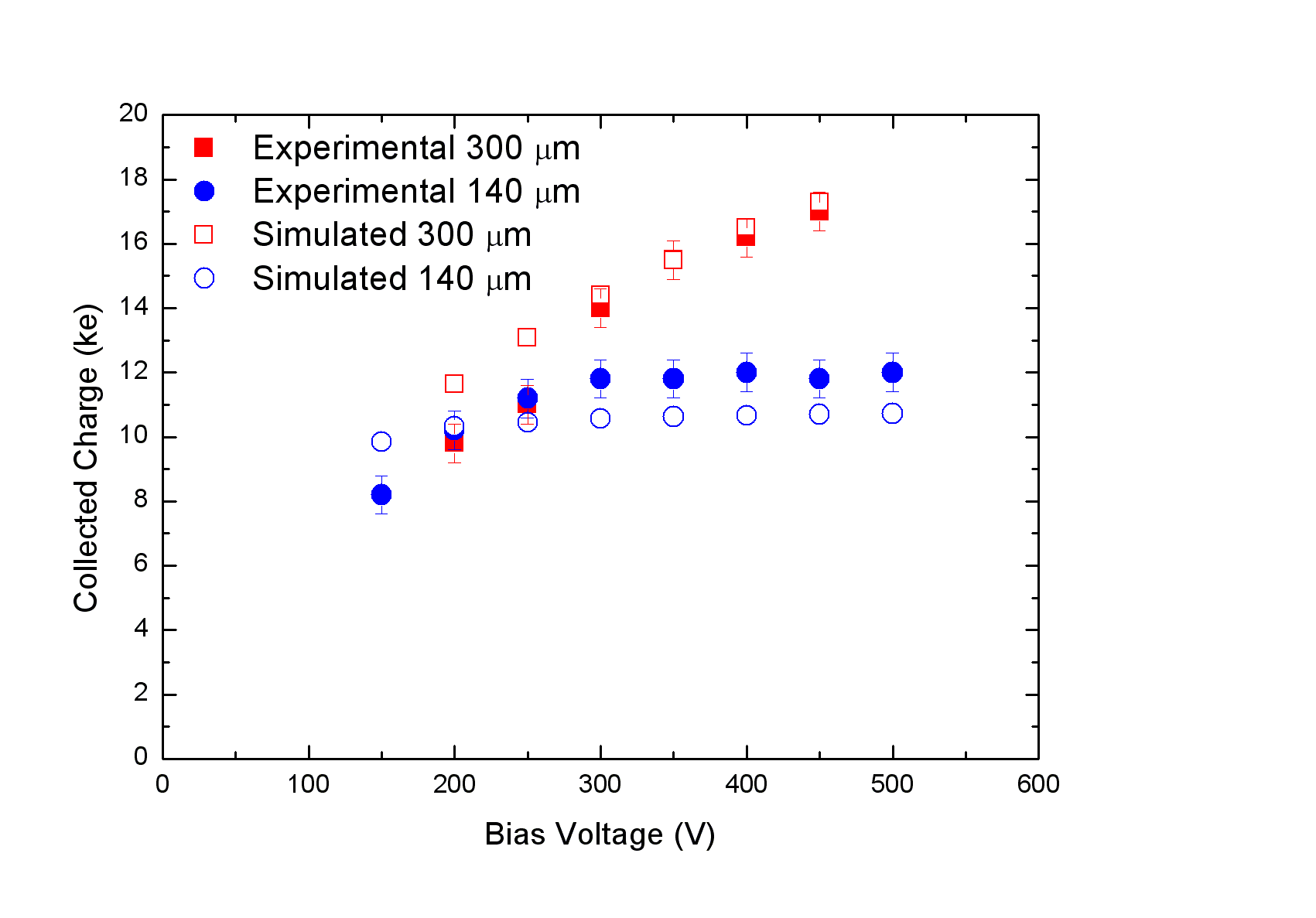}
\caption{Comparison of measured CCE for thin and thick n-in-p detectors with transient simulations.}
\label{fig15}
\end{figure}

We have studied the charge collection through transient simulations by shooting a Minimum Ionizing particle~(MIP) orthogonally on one of the junction side strips and observing the charge collection on the corresponding ohmic side strip. Figures~\ref{fig13}(a) and ~\ref{fig13}(b) show the transient signals observed on the p-side of DSSD exposed to a fluence of 3.93$\times$10$^{12}$~n$_{eq}$cm$^{-2}$ and 20.60$\times$10$^{12}$~n$_{eq}$cm$^{-2}$ respectively. One can infer from these plots that the charge collection time in either case can be improved by operating the sensors at high bias voltage since in either case less mobile holes are being collected which can be confirmed from the polarity of the signal. Figure~\ref{fig14}(a) and ~\ref{fig14}(b) show the transient signals observed on the n-side of DSSD exposed to a fluence of 3.93$\times$10$^{12}$~ n$_{eq}$cm$^{-2}$ and 20.60$\times$10$^{12}$~n$_{eq}$cm$^{-2}$ respectively. It can be noted that for this case, charge collection is much faster as n-side is collecting electrons. These transient signals have been integrated over time to extract the collected charge.\\
In order to validate the transient simulations, we have reproduced the experimentally observed CCE for thin (140~$\mu$m) and thick (300~$\mu$m) n-in-p detectors irradiated to a fluence of 5.0$\times$10$^{14}$~n$_{eq}$cm$^{-2}$. The structure and substrate doping of the simulated device matched with those tested in Ref.~\cite{pp}. The comparison of the simulated CCE with the measurements can be seen in Figure~\ref{fig15}. The simulation results agree with measurements within 10~\% error thus validating the transient simulations. 

\begin{figure}[htb!]
\vspace*{4mm}
\centering
\includegraphics*[height=60mm]{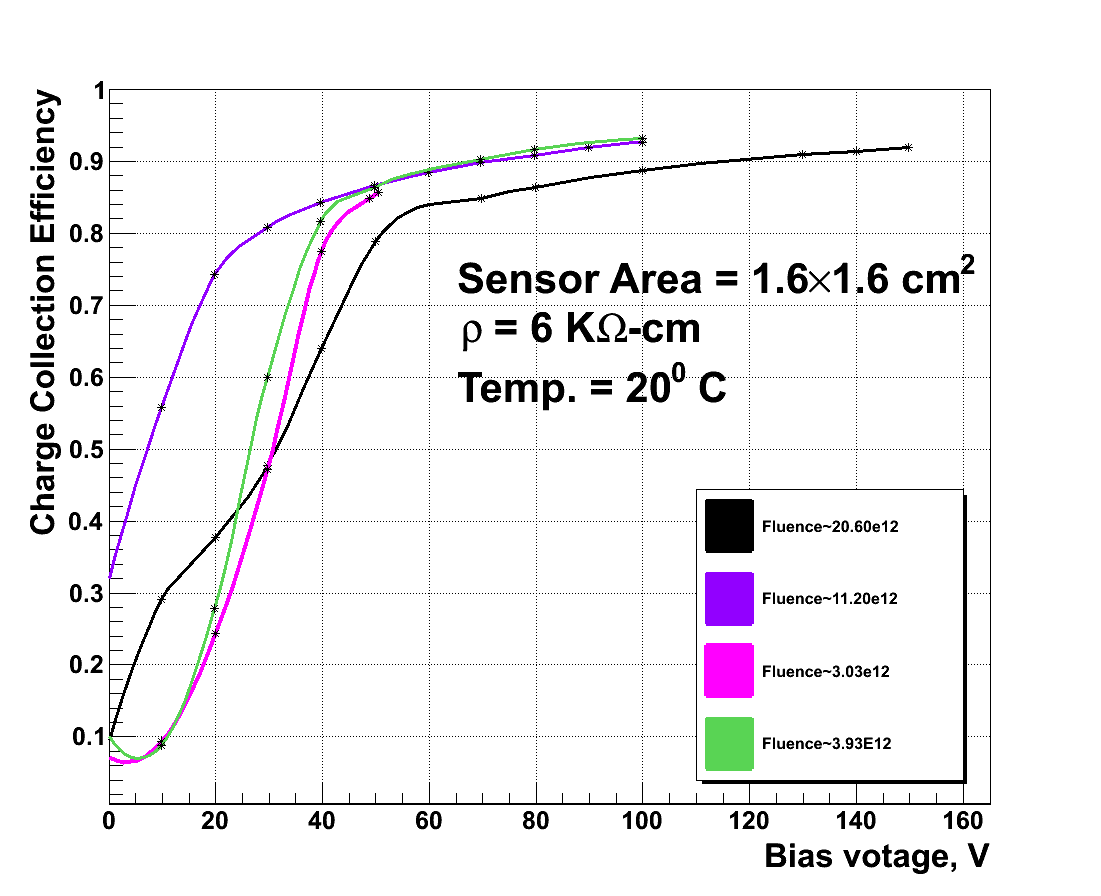}
\caption{CCE versus bias voltage for irradiated DSSDs equipped with P-stop isolation.}
\label{fig16}
\end{figure}

\begin{figure}[htb!]
\vspace*{4mm}
\centering
\includegraphics*[height=60mm]{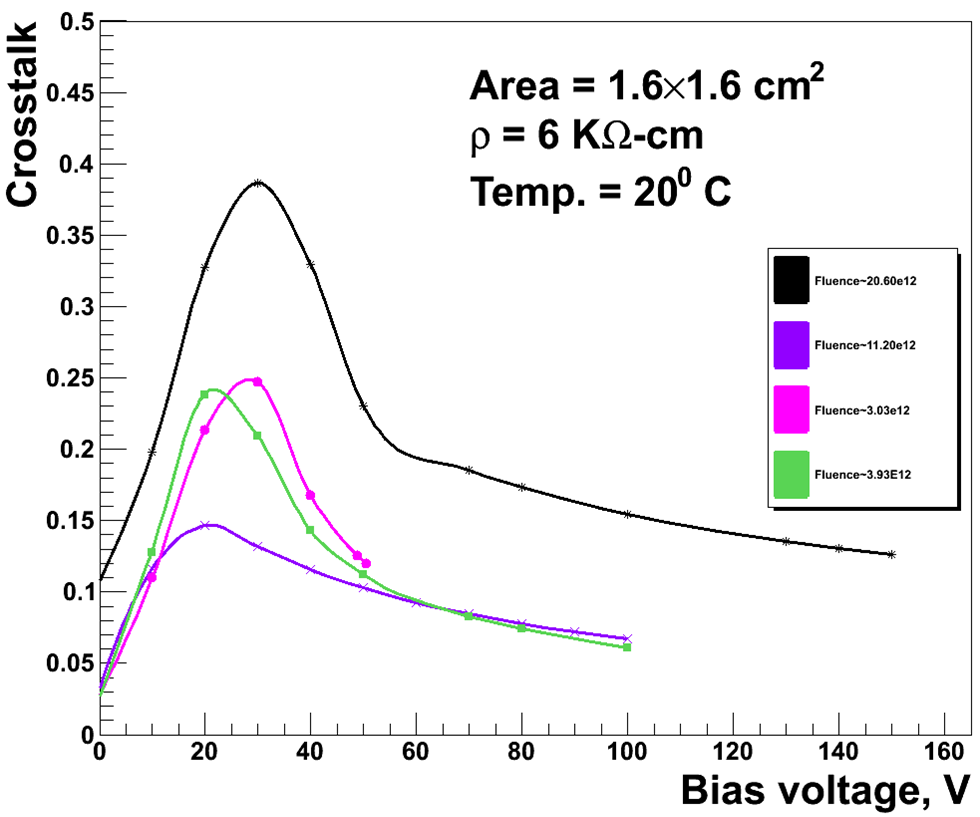}
\caption{Cross-talk versus bias voltage for irradiated DSSDs equipped with P-stop isolation.}
\label{fig17}
\end{figure}

\begin{figure}[htb!]
\vspace*{4mm}
\centering
\includegraphics*[height=60mm]{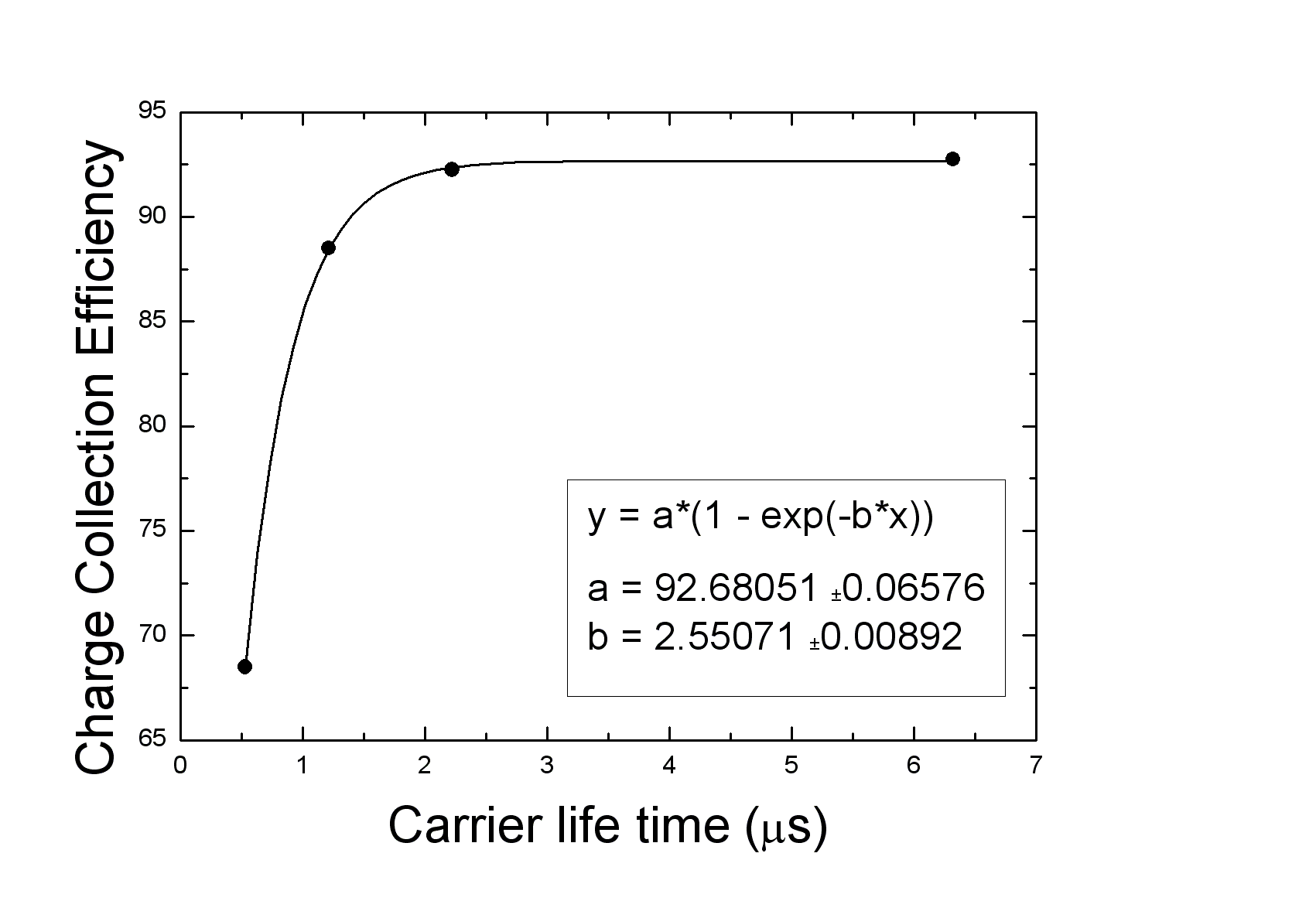}
\caption{Modelling of dependence of CCE on carrier life time.}
\label{fig18}
\end{figure}

Figure~\ref{fig16} shows the variation of collected charge versus V$_{bias}$ for the same fluences as used for irradiation. The charge collection follows the same variation as measured R$_{int}$ does with V$_{bias}$. Figure~\ref{fig17} shows the charge pickup (crosstalk) by neighbouring strip on the ohmic side as a function of V$_{bias}$ . As expected, the crosstalk initially increases with V$_{bias}$ for all the fluences since the depletion region gets wider, and so the amount of charge sharing between the neighbouring strips will increase. On the other hand, increasing the bias will also improve the drift velocity, which tends to reduce charge sharing. So, as the detector nears full depletion, this second effect becomes more important, and the crosstalk slowly decreases. It becomes clear from Figures~\ref{fig16} and ~\ref{fig17} that the main impact of irradiation is the deterioration of CCE. We have tried to extract a mathematical model for the dependence of CCE on the carrier life time as can be seen in Figure~\ref{fig18}.

\subsection{Design optimization of isolation techniques in DSSDs}
\label{}

\begin{figure}[htb!]
\vspace*{4mm}
\centering
\includegraphics*[height=50mm]{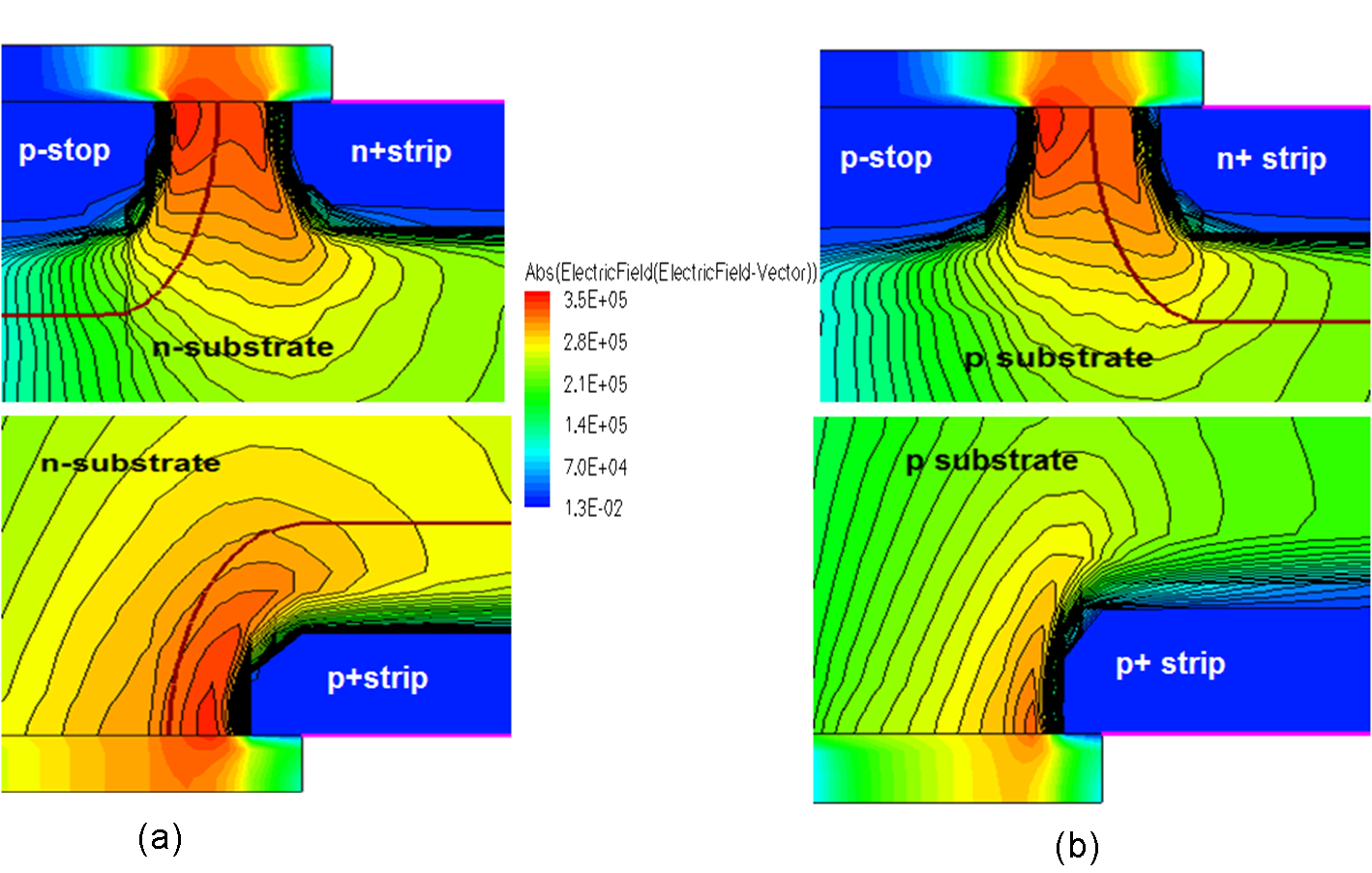}
\caption{Simulated electric field distribution on n and p side of the DSSD at breakdown irradiated to a fluence of (a) 3.93$\times$10$^{12}$~n$_{eq}$cm$^{-2}$ (before type-inversion) and (b) 20.60$\times$10$^{12}$~n$_{eq}$cm$^{-2}$ (beyond type-inversion).}
\label{fig19}
\end{figure}

\begin{figure}[htb!]
\vspace*{4mm}
\centering
\includegraphics*[height=50mm]{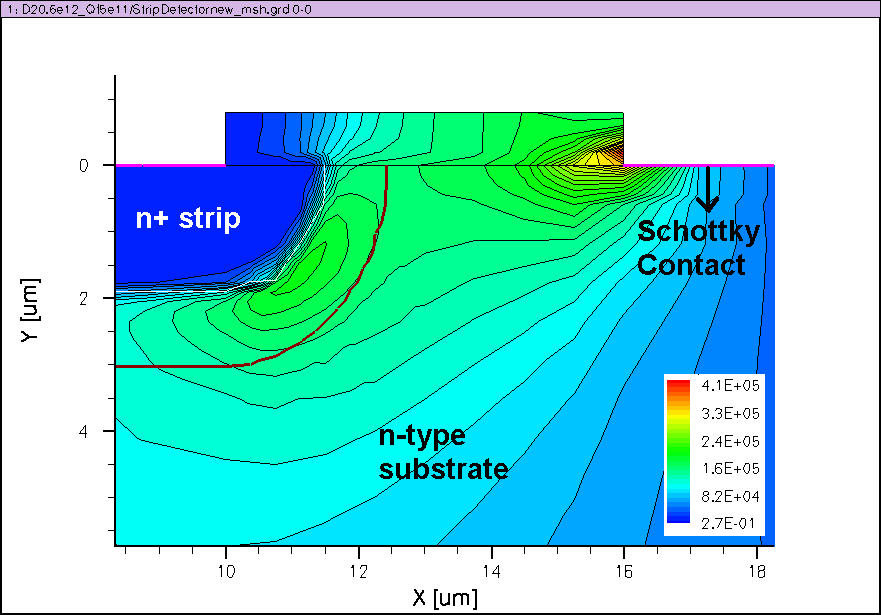}
\caption{Simulated electric field distribution at breakdown for DSSD equipped with schottky barrier.}
\label{fig20}
\end{figure}

Table~\ref{table:table4} shows the expected breakdown voltage (V$_{bd}$) for DSSDs exposed to a low fluence (before type-inversion) and high fluence (after type-inversion) having three isolation techniques namely P-stop, P-spray and Schottky barrier. The CBM plans to use floating electronics ~\cite{uu}, hence for V$_{bd}$ simulations, DC coupled DSSDs have been considered. One can notice that in all the three cases, the V$_{bd}$ deteriorates with fluence. For DSSDs having conventional isolation techniques, this happens since before type-inversion the electric field is distributed on either sides while after type-inversion the high electric field exists only on the n- side.  This can be seen from the Figure~\ref{fig19} . However in the case of DSSDs having Schottky barrier, the critical electric field responsible for breakdown occurs in the oxide between n-strips and schottky contact as can be seen from Figure~\ref{fig20}. Hence the V$_{bd}$ deteriorates with fluence since there is an increase in surface oxide charge with fluence. One can infer from Table~\ref{table:table4} that in terms of breakdown performance, the Schottky barrier is the best choice.

\begin{table*}[htb!] \small
\caption{Comparison of conventional isolation techniques with schottky barrier isolation in terms of static and dynamic parameters.}
\centering
\begin{tabular}{|c|c|c|c|c|c|c|}
\hline
Isolation & Fluence & V$_{bd}$ & \multicolumn{2}{c|} {C$_{int}$ (pF cm$^{-1}$)} & R$_{int}$@ 80V& CCE @ 80V \\ [0.5ex]
\cline{4-5}
Technique& (n$_{eq}$ cm$^{-2}$) & (V) & n-side & p-side & (M$\Omega)$ & ~\%  \\ [0.5ex]
\hline 
\multirow {2} {*} {P-spray} & 3.93$\times$10$^{12}$ & 524  & 2.6 & 1.7 & 1250  & 93\\[0.5ex]
&20.60$\times$10$^{12}$ & 450  & 2.7 & 2.1 & 104 & 86.25\\ \hline
\multirow {2} {*} {P-stop} & 3.93$\times$10$^{12}$ & 860  & 2.1 & 1.7 & 1136 & 91.25\\[0.5ex]
&20.60$\times$10$^{12}$& 750  & 2.29 & 2.1 & 125 & 86.25 \\\hline
\multirow {2} {*} {Schotkky} & 3.93$\times$10$^{12}$  & 1450 & 2.05 & 1.7 & 19.64  & 79 \\[0.5ex]
Barrier &20.60$\times$10$^{12}$ & 1350  & 1.8 & 4.0 & 8 & 77.5 \\
\hline
\end{tabular}
\label{table:table4}
\end{table*}

\begin{table*}[htb!] \small
\caption{Optimization of modulated p-spray}
\centering
\begin{tabular}{|c|c|c|c|c|c|c|c|c|c|c|c|c|}
\hline\hline
p-spray dose (cm$^{-3}$) & \multicolumn{3}{c|}{1$\times$10$^{15}$ (very low dose)} &  \multicolumn{3}{c|}{4$\times$10$^{16}$ (low dose)}& \multicolumn{3}{c|}{8$\times$10$^{16}$ (medium dose)}& \multicolumn{3}{c|}{12$\times$10$^{16}$ (high dose)}\\
\hline
p-stop width ($\mu$m)& V$_{bd}$(V) & \multicolumn{2}{c|} {C$_{int}$(pF cm$^{-1}$)} & V$_{bd}$(V) & \multicolumn{2}{c|} {C$_{int}$(pF cm$^{-1}$)} &V$_{bd}$ (V)& \multicolumn{2}{c|} {C$_{int}$(pF cm$^{-1}$)} &V$_{bd}$ (V)& \multicolumn{2}{c|} {C$_{int}$(pF cm$^{-1}$)} \\ [0.5ex]

\cline{3-4}\cline{6-7}\cline{9-10}\cline{12-13}

&  & n-side & p-side & & n-side & p-side& & n-side & p-side& & n-side & p-side \\
\hline
5 &  1125 &1.56 &1.75  & 490 & 2.405 &1.75 & 210 & 2.46&1.76  & 161 & 2.47&1.77   \\\hline
10 & 1125 &1.5 &1.75  & 480 & 2.4&1.75  & 205 & 2.46&1.76  & 160 & 2.47&1.77  \\\hline
15 & 1150 &1.6&1.75  & 488 & 2.37 &1.75 & 205 & 2.43&1.76  & 160 & 2.47 &1.77 \\\hline
20 & 650 &2.09 &1.75 & 450 & 2.51 &1.75 & 205 & 2.42&1.76  & 160 & 2.47 &1.77 \\ [1ex]
\hline
\end{tabular}
\label{table:table5}
\end{table*}

\begin{table}[htb!] \small
\caption{Comparison between p-stop, p-spray and optimized modulated p-spray at low and high fluence.}
\centering
\begin{tabular}{|c|c|c|c|c|}
\hline
Isolation & Fluence & V$_{bd}$ & C$_{int}$  & CCE \\ [0.5ex]
Technique& (n$_{eq}$ cm$^{-2}$) & (V) & (pF cm$^{-1}$)&  ~\%  \\ [0.5ex]
\hline 
\multirow {2} {*} {P-stop} & 2$\times$10$^{13}$ & 800  & 2.08 & 93.15 \\[0.5ex]
&1$\times$10$^{14}$ & 610 & 2.09  & 88.87  \\ \hline
\multirow {2} {*} {P-spray} & 2$\times$10$^{13}$ & 513 & 2.56 & 93.17  \\[0.5ex]
&1$\times$10$^{14}$& 495 & 2.44 & 89  \\\hline
\multirow {2} {*} {Optimized} & 2$\times$10$^{13}$  & 1600 & 1.58 & 93.22  \\[0.5ex]
Modulated P-spray &1$\times$10$^{14}$ & 1150 & 1.60 & 89  \\
\hline
\end{tabular}
\label{table:table6}
\end{table}

Interstrip capacitance (C$_{int}$) has been simulated for DSSDs having conventional isolation techniques and also for DSSDs equipped with Schottky barrier both for low and high fluences as can be seen from Table~\ref{table:table4}. One can notice that the C$_{int}$ on the p-side is lower as compared with n-side for DSSDs equipped with P-stop and P-spray. This again can be explained based on e$^{-}$ accumulation layer ~\cite{oo} which helps in coupling the n-strips while doing an opposite effect between the p-strips. Also one can notice that with increasing fluence, the C$_{int}$ on the p-side increases in larger proportion as compared with n-side. This could be explained since the silicon bulk inverts to P-type at higher fluence, which helps to overcome the inhibition of hole density by the electron accumulation layer between the p-strips. In the case of DSSDs having schottky barrier, a steep rise in the value of C$_{int}$ is observed after type-inversion on the p-side. This has also been observed during measurements. Since the C$_{int}$  on the p-side of DSSDs having schottky barrier is too high at high fluence, this is not a good choice for isolation technique in terms of capacitive noise.

\begin{figure}[htb!]
\vspace*{4mm}
\centering
\includegraphics*[height=60mm]{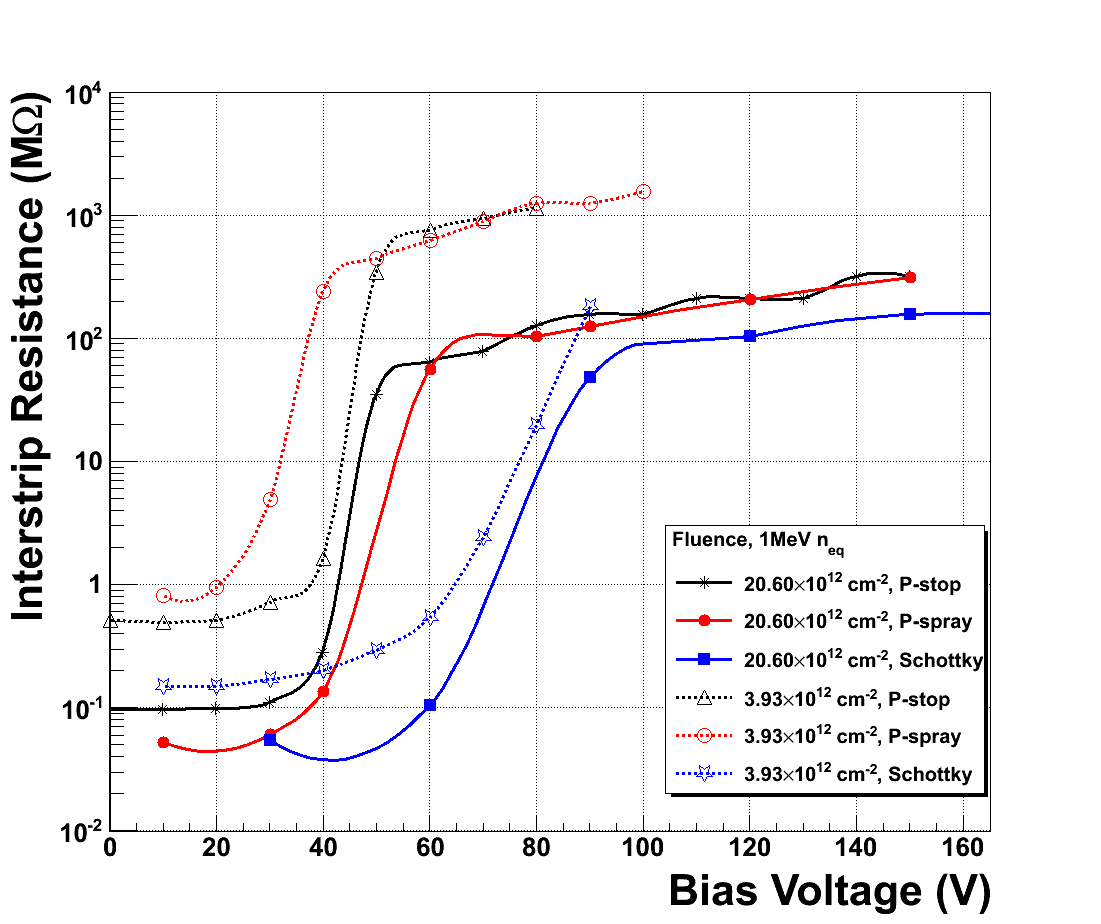}
\caption{Comparison of R$_{int}$ vs. V$_{bias}$ for DSSDs equipped with P-stop, P-spray and Schottky barrier at low and high fluences.}
\label{fig21}
\end{figure}

\begin{figure}[htb!]
\vspace*{4mm}
\centering
\includegraphics*[height=60mm]{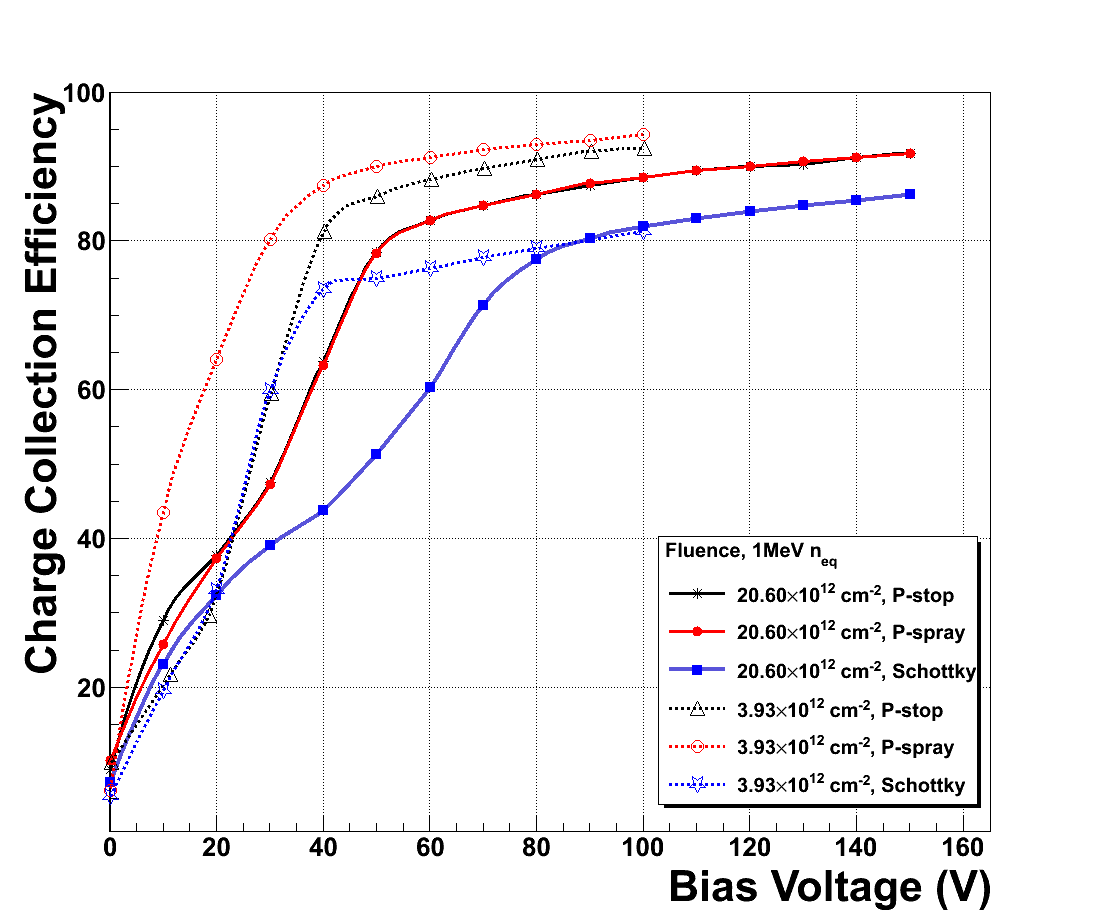}
\caption{Comparison of CCE vs. V$_{bias}$ for DSSDs equipped with P-stop, P-spray and Schottky barrier at low and high fluences.}
\label{fig22}
\end{figure}

Figure~\ref{fig21} shows the variation of R$_{int}$ versus V$_{bias}$ for DSSDs equipped with P-stop, P-spray and Schottky barrier at low and high fluences. One can observe that for low fluence, P-spray gives better isolation as compared with P-stop. The operating voltage with P-spray isolation technique is reached at 35~V while with P-stop, the operating voltage is at 45~V. The R$_{int}$ for DSSD with schottky barrier is comparatively very low and its operating voltage is reached at 90~V. For high fluence, the R$_{int}$ of P-spray and P-stop are almost same and both reach their operating voltage at around 70V. Again for DSSD having schottky barrier, R$_{int}$ is throughout low and the operating voltage is reached at 100~V. As discussed in Section 3.3, higher operating voltage means higher power consumption which could lead to thermal runaway at high fluences. Figure~\ref{fig22} shows the variation of CCE versus V$_{bias}$ for DSSDs having conventional isolation techniques and schottky barrier. Again the CCE mimics the variation of R$_{int}$ with V$_{bias}$. For low fluence, P-spray gives better CCE as compared with P-stop while Schottky barrier gives the worst performance in terms of CCE. As in the case of R$_{int}$, at high fluence, the CCE of P-stop and P-spray is same while for schottky barrier, the CCE is very poor. Hence in terms of power consumption and CCE, it seems not optimal to use Schottky barrier as an effective isolation technique. 

An optimization study has been done for modulated P-spray in terms of P-spray dose and P-stop width for the maximum expected fluence of 1$\times$10$^{14}$~ n$_{eq}$cm$^{-2}$. It is known that narrow P-stop improves the V$_{bd}$ but deteriorates the C$_{int}$ ~\cite{qq,rr} while in the case of P-spray, higher dose implies premature breakdown and higher C$_{int}$ for the p-type substrate. We have taken various combinations of P-stop width and P-spray dose and extracted the V$_{bd}$ and C$_{int}$. As can be seen from Table~\ref{table:table5}, the C$_{int}$ increases and V$_{bd}$ decreases with P-spray dose for a fixed P-stop width. Also, one can observe from Table~\ref{table:table5} that for a fixed P-spray dose, the V$_{bd}$ initially increases with P-stop width till 15~$\mu$m but deteriorates afterwards. This effect is more prominent for very low and low P-spray dose. From this table, it becomes clear that the best design criteria is to combine very low P-spray dose (1$\times$10$^{15}$~cm$^{-3}$) with 15~$\mu$m P-stop width, referred to as Optimized Modulated P-spray in Table~\ref{table:table6} .  Finally a comparison of P-stop, P-spray and Optimized Modulated P-spray for 1$^{st}$ year of CBM run fluence (2$\times$10$^{13}$~ n$_{eq}$cm$^{-2}$) and the maximum fluence expected at the end of five years of CBM run (1$\times$10$^{14}$~ n$_{eq}$cm$^{-2}$) has been shown in Table~\ref{table:table6}. It is clear from this table that optimized modulated P-spray is the best choice for isolation technique in terms of V$_{bd}$, C$_{int}$ and CCE.

\section{Conclusions}
\label{}
This paper reports on the ongoing R\&D effort to develop radiation hard low noise radiation hard DSSDs for the upcoming CBM experiment at FAIR. An optimization of isolation techniques has been done to maximize CCE and V$_{bd}$ while minimizing the total ENC. Transient simulations has proven to be an effective tool to determine the operating voltage and to predict the expected CCE.

\section{Acknowledgement}
\label{}
The authors would like to thank Dr. M. Merkin and Dr. D. Karmanov from SINP, Moscow State University, Russia for irradiating CBM prototypes. We would also like to thank Dr. David Pennicard from DESY, Germany and Dr. Gemma Kerr from Microelectronics support centre, Rutherford Appleton Laboratory, UK for useful discussions and suggestions.



\begin{thebibliography}{12}
\bibitem{aa} 
http://www.fair-center.eu/en/fair-users/experiments/cbm.html

\bibitem{n-XYTER}
A. S. Brogna et al.,"N-XYTER, a CMOS read-out ASIC for high resolution time and amplitude measurements on high rate multi-channel
counting mode neutron detectors,"NIM A, vol. 568, 2006, pp. 301-308.

\bibitem{bb}
G. Barichello et. al.,"Performance of long modules of Silicon Microstrip detectors," NIM A, vol. 413, 1998, pp. 17-30.

\bibitem{cc}
C. Bozzi, "Signal-to-Noise evaluations for the CMS Silicon Microstrip Detectors," CMS note 1997/026.

\bibitem{dd}
http://pdg.lbl.gov/2010/reviews/rpp2010-rev-particle-detectors-accel.pdf.

\bibitem{ee} 
[1]R. Wunstorf, "Radiation hardness of silicon detectors:current status," IEEE Trans. Nucl. Sci., vol. 44, 1997, pp.806.

\bibitem{ff} 
[1]S. Chatterji et. al., "Simulation study of irradiated Si sensors equipped with metal- overhang for applications in LHC environment," vol.51 (2), 2004, pp.298-312.

\bibitem{gg} 
http://www.pfk.ff.vu.lt/lectures/funkc\_dariniai/diod/schottky.htm

\bibitem{zz} 
http://ecourses.vtu.ac.in/nptel/courses/Webcourse-contents/IIT-Delhi/Semiconductor$\%$20Devices/metal\_semi/lec1.htm

\bibitem{yy} 
http://www.synopsys.com/home.aspx

\bibitem{hh} 
"RD 48 Status Report", CERN/LHC 2000-009, Dec.1999.

\bibitem{ii} 
[1]M. Moll et. al., " Investigation on the improved radiation hardness of silicon detectors with high oxygen concentration," NIM A, vol.439, 2000, pp.282-292.

\bibitem{jj} 
H. W. Kraner, Z.Li and K.U. Posnecker, "Fast neutron damage in silicon detectors," NIM A, vol.225, 1984, pp.615.

\bibitem{kk} 
J. G. Fossum, R. P. Mertens, D. S. Lee, and J. F. Nijs, "Carrier Recombination and Lifetime in Highly Doped Silicon," Solid-State Electron., vol.26 (6), 1983, pp. 569-576.

\bibitem{ll}
V. Eremin et. al., "Effect of radiation induced deep level traps on Si detector performance," NIM A, vol.476 (3), 2002, pp.537-549.

\bibitem{xx}
RD48 3rd Status Report, CERN/LHCC 2000-009.

\bibitem{ww}
G.Lindstrom et.al., NIM A 426 (1999)1.

\bibitem{mm}
W. Shockley and W. T. Read, "Statistics of  the Recombinations of Holes and Electrons,"  Phys.Rev., vol.87 (5), 1952, pp.835-842.

\bibitem{nn}
M. Petasecca, F. Moscatelli, D. Passeri, G.U. Pignatel, "Numerical Simulation of radiation Damage Effects in p-Type and n-Type FZ Silicon Detectors ," IEEE Trans. Nucl. Sci. NS-53 (5), 2006, pp.2971.

\bibitem{oo}
N. Tamura et. al., "Radiation effects of double-sided silicon strip sensors," NIM A, vol. 342(1), 1994, pp. 131-136.

\bibitem{pp}
G.Casse et.al., "Charge Collection Efficiency Measurements for segmented Silicon Detectors Irradiated to 1$\times$10$^{16}$n~cm$^{-2}$," IEEE Trans. Nucl. Sci., vol.55 (3), 2008, pp.1695.

\bibitem{uu}
T.Kawasaki, "The Belle Silicon Vertex Detectors," NIM A 494, 2002, pp.94-101.

\bibitem{qq}
C.Piemonte, "Device Simulations of Isolation Techniques for Silicon Microstrip Detectors Made on p-type Substrates," IEEE Trans. Nucl. Sci. vol.53 (3), 2006, pp.1694.

\bibitem{rr}
G.-F.Dalla Betta, "Surface Effects and Breakdown Voltage," MC-PAD Training Event, Ljubljana, 27 September, 2010.



\end{thebibliography}
\end{document}